\documentstyle[12pt,fleqn]{article}
\newcommand{\sbody}[2]{{\textstyle\frac{#1}{#2}}}

\begin{document}
\begin{center}
\large\bf{Direct Single-Instanton Contributions}\\
\large\bf{to Finite-Energy Sum Rules}
\end{center}
\begin{center}
A.S. Deakin, V. Elias, Ying \vspace{-.5cm}Xue\\
Department of Applied \vspace{-.5cm}Mathematics\\
University of Western \vspace{-.5cm}Ontario\\
London, Ontario  N6A \vspace{-.5cm}5B7\\
Canada\\
N.H. \vspace{-.5cm}Fuchs \\
Department of \vspace{-.5cm} Physics\\
Purdue \vspace{-.5cm} University\\
West Lafayette, IN  \vspace{-.5cm}47907\\
U.S.A.\\
Fang Shi and T.G. \vspace{-.5cm}Steele\\
Department of Physics and Engineering \vspace{-.5cm}Physics\\
University of \vspace{-.5cm}Saskatchewan\\
Saskatoon, Saskatchewan  S7N \vspace{-.5cm}5C6\\
Canada
\end{center}\vspace{.1cm}
\begin{center}
{\bf{Abstract}}\end{center}
\hspace{.5cm} 
Instanton contributions to pseudoscalar finite-energy sum rules are 
extracted from the explicit
single-instanton contribution to the pseudoscalar Laplace sum rule in 
the instanton liquid model.
\eject
Finite energy sum rules in the pseudoscalar meson channels have been 
used by a number of
researchers to obtain bounds on quark masses [1,2,3].  Substantial 
higher-order perturbative
contributions to the pseudoscalar correlation function are both known
[2,3,4] and controllable [5]. 
As emphasized in ref. 3, however, such calculations are understood to 
be subject to serious
uncertainties from direct instanton contributions [6], which have not 
been sufficiently 
well-understood to be incorporated into even the most recent finite 
energy sum rule calculations [2].
In this note, 
we use the known contribution to Laplace
sum rules in the instanton liquid model [7]
to extract the direct 
single-instanton contribution 
to finite-energy sum rules: 
\begin{eqnarray}R_0^{\,p} (s)& = &\frac{1}{\pi} \; \int_0^\infty \; Im \left[ 
\left(\pi^p(t)\right)_{inst} \right] e^{-
st}dt
\nonumber \\
&=& \left(\frac{4\pi^2n_c\rho^2}{3m_*^2}\right)
\frac{3\rho^2}{8\pi^2s^3} \; e^{-\rho^2/2s} \left[K_0
\left(\frac{\rho^2}{2s}\right) + K_1
\left(\frac{\rho^2}{2s} \right) \right]
\nonumber
\\
&=& \frac{3\rho^2}{8\pi^2s^3} \; e^{-\rho^2/2s} \left[K_0
\left(\frac{\rho^2}{2s}\right) + K_1
\left(\frac{\rho^2}{2s} \right) \right].\label{(1)}
\end{eqnarray}
In (\ref{(1)}), $1/\rho\approx 600\,{\rm MeV}$ is the instanton size,  $s$ is 
the Borel parameter ($s
\equiv
1/M^2$) and $\pi^p(q^2)$ denotes the correlator of appropriate light-quark 
pseudoscalar currents $i\bar q \gamma_5 q$.  
In the instanton liquid model
the quantity $n_c$ parametrizes the 
instanton density and $m_*$ is the self-consistent dynamical mass. 

The finite energy sum rules we wish to obtain are
\begin{equation}F_k^{\;p}(s_0) \equiv \frac{1}{\pi} \; \int_0^{s_0} \; 
Im\left[\left(\pi^p (t)\right)_{inst}
\right]t^k
dt.\label{(2)}\end{equation}
To evaluate the contributions to (\ref{(2)}) in the instanton liquid model,
recall that $R_0^p(s)$ in (\ref{(1)}) is itself a Laplace transform:
\begin{eqnarray}
& & R_0^{\;p}(s) = {\cal{L}} \left[ \frac{1}{\pi} \, Im 
\left(\pi^p (t) \right)_{inst}
\right],\label{(3)}
\\
& &{\cal{L}} [f(t)] \equiv \int_0^\infty\, f(t) e^{-st} dt.\label{(4)}
\end{eqnarray}
From (\ref{(2)}) and (\ref{(3)}) we see that
\begin{equation}
\frac{d}{dt} \; F_k^p (t) = {\cal{L}}^{-1} \left[ R_0^p 
(s)\right]t^k.\label{(5)}
\end{equation}
Upon taking the Laplace transform of both sides of (\ref{(5)}) and noting 
from (\ref{(2)}) that $F_k^p(0) = 0$,
we obtain
\begin{equation}
F_k^p(t) = {\cal{L}}^{-1} \left[ \frac{1}{s} \left(-\frac{d}{ds} 
\right)^k
R_0^p(s)\right]\label{(6)}
\end{equation}
An explicit expression for $F_k^p (t)$ can
be obtained from the identity [8]
\begin{eqnarray}
\frac{1}{2s}\; e^{-1/2s} K_0(1/2s) &= & -\pi \int_0^\infty \;
J_0(x)Y_0(x)e^{-sx^2} x\; dx\nonumber\\
&=& {\cal{L}} \left[ - \frac{\pi}{2} \; J_0 (\sqrt{t}) 
Y_0
(\sqrt{t})\right]\; . \label{(7)}
\end{eqnarray}
We differentiate both sides of (\ref{(7)}) with respect to $s$, noting that
$K_0^\prime (z) = - K_1(z)$ and that ${\displaystyle{\frac{d}{ds}}}\,
{\cal{L}}[f(t)] = {\cal{L}}[-t f(t)]$, in order to obtain the
relation
\begin{eqnarray}
& &H_0 (s)  \equiv  \frac{1}{(2s)^3} \, e^{-1/2s} \left[ K_0 (1/2s) +
K_1(1/2s) \right]=  {\cal{L}}[h(t)],\label{(8a)}
\\
h(t) &=& \frac{\pi}{4} \,t\, J_0 (\sqrt{t}) Y_0 (\sqrt{t}) + 
{\cal{L}}^{-1} \left[ \frac{1}{2s} \,
{\cal{L}}\left[ - \frac{\pi}{2} \, J_0 (\sqrt{t}) Y_0 (\sqrt{t}) 
\right] \right]\nonumber\\
&=&  \frac{\pi}{4} \,t\, J_0 (\sqrt{t}) Y_0 (\sqrt{t}) - 
\frac{\pi}{4} \, \int^{t}_0 \, J_0 (\sqrt{w})
Y_0 (\sqrt{w}) dw, \label{(8b)}
\end{eqnarray}
where the integral in the final line above 
is a convolution of $J_0 (\sqrt{t}) Y_0 
(\sqrt{t})$ and 
$1/2 = {\cal{L}}^{-1}(1/2s)$.  Comparing the top line of (\ref{(8a)}) with 
(\ref{(1)}), we see that
\begin{equation}
R_0^p (s) = \frac{3}{\pi^2 \rho^4}\, H_0(s/\rho^2)\label{(9)}
\end{equation}
Using the rescaling
relation $G(s/\rho^2) = \rho^2 {\cal{L}} [g(\rho^2 t)]$ for $G(s) =
{\cal{L}} [g(t)]$, one can easily show via (\ref{(6)}) and (\ref{(9)}) that
\begin{equation}F_k^p (t) = \frac{3}{\pi^2\rho^{4+2k}} \, \phi_k 
(\rho^2t)\label{(10a)}
\end{equation}
where       
\begin{equation}
\phi_k(t) = {\cal{L}}^{-1} \left[ \frac{1}{s} \left(- \frac{d}{ds} 
\right)^k H_0
(s)\right]
 = \int_0^t \, \tau^k h(\tau)d\tau. \label{(10b)}
 \end{equation}
We find from substitution of (\ref{(8b)}) into (\ref{(10b)}) that
\begin{eqnarray}
\phi_k (t) &=& \frac{\pi}{4} \, \int_0^t \, d\tau \, 
\tau^k \left[ \tau J_0 (\sqrt{\tau}) Y_0
(\sqrt{\tau})
- \int_0^\tau \, dw \, J_0 (\sqrt{w}) Y_0 (\sqrt{w}) 
\right]\nonumber
\\
& = &\frac{\pi}{4(k+1)} \, \int_0^t \left[ 
(k+2)\tau^{k+1} - t^{k+1} \right] J_0 (\sqrt{\tau}) Y_0
(\sqrt{\tau})d\tau.\label{(11)}
\end{eqnarray}
Substitution of (\ref{(11)}) into (\ref{(10a)}) yields 
a closed-form expression for 
the instanton contribution (\ref{(2)})
to finite energy sum rules:
\begin{equation}
F_k^p(s_0) = \frac{3}{4\pi(k+1)} \int_0^{s_0} \, \left[ 
(k+2)w^{k+1} - s_0^{k+1} \right]J_0
(\rho \sqrt{w}) Y_0 (\rho\sqrt{w}) dw.\label{(12)}
\end{equation}
The appearance of the explicit $s_0^{k+1}$ term in (\ref{(12)}) reminiscent
of perturbative contributions, 
raises the concern
that the instanton and perturbative contributions might be comparable.  
A simplification of (\ref{(12)}) addresses this question.
Applying a change of variables in (\ref{(12)}), using the identity
\begin{equation}
\int xJ_0(x) Y_0(x)\, dx=\frac{1}{2}x^2\left[ J_0(x)Y_0(x)+J_1(x)Y_1(x)\right]
\label{tom1}
\end{equation}
and performing an  integration by parts results in the expression
\begin{equation}
F_k^p\left(s_0\right)=-\frac{3}{4\pi}\int\limits_0^{s_0}
w^{k+1}J_1\left(\rho\sqrt{w}\right) Y_1\left(\rho\sqrt{w}\right)\, dw
\label{tom2}
\end{equation}
which, by comparing the integrands, is easily seen to be smaller 
than the leading perturbative contribution.
From comparison of (\ref{tom2}) and (\ref{(2)}) it is also possible 
to make the identification
\begin{equation}
\frac{1}{\pi}Im\,\left[\pi^p(w)\right]_{inst}=-\frac{3}{4\pi}
wJ_1\left(\rho\sqrt{w}\right) Y_1\left(\rho\sqrt{w}\right)
\label{tom3}
\end{equation}

Approximate expressions for the inverse Laplace transforms (\ref{(6)}) in
terms of elementary trigonometric functions may be obtained via
asymptotic expansion methods in the complex plane.  We rewrite (\ref{(6)}) as
follows:
\begin{equation}
F_k^p(t) = \frac{1}{2\pi i} \; \int_C \left[ \frac{1}{s} \left(- 
\frac{d}{ds}\right)^k
R_0^{\;p} (s)
\right]e^{st}
ds,\label{(13)}
\end{equation}
with the contour $C$ in the complex $s$ plane [Fig. 1] being a 
vertical line  on which $Re(s)$
is a positive constant.  We can distort $C$ as indicated in Fig. 2.  
The arc
contributions $C_1$ and $C_2$ vanish, because as $|s| \rightarrow 
\infty$
\begin{equation}
R_0^{\;p} (s) \longrightarrow \frac{1}{|s|^2}\; , \label{(14)}
\end{equation}
as is evident from (\ref{(1)}).  Consequently, $F_k^{\,p}(t)$ can be 
expressed as an integral around the
Hankel loop contour $L$ given in Fig. 2.

To proceed further, we make use of the asymptotic expansion [9]
\begin{eqnarray}
& &K_0(z) + K_1(z) \sim \left( \frac{\pi}{2z} \right)^{\sbody12} e^{-z}
\sum_{n=0}^\infty \; a_n z^{-
n},\label{(15a)}\\
& &a_0 = 2,\;\; a_1 = \sbody14,\;\; a_2 = -\sbody{3}{64},\;\; a_3 = 
\sbody{15}{512},\;\; \ldots
\label{(15b)}
\end{eqnarray}
in order to obtain the following result:
\begin{equation}F_k^p(t) \sim \sum_{n=0}^\infty \; b_n\; 
\frac{1}{2\pi i} \int_L \; 
\frac{e^{st}}{s} \left[ \left(
-
\frac{d}{ds}\right)^k \left[e^{-\rho^2/s} s^{n-3+1/2} \right]\; 
\right] ds
\label{(16a)}
\end{equation}
where
\begin{equation}
b_n \equiv \frac{3\rho^{1-2n}}{8\pi^{3/2}}\; 2^n a_n \; 
.\label{(16b)}
\end{equation}
The integrals in (\ref{(16a)}) can be evaluated through explicit use of 
Schl\"afli's integral
[10] over the
Hankel contour $L$:
\begin{equation}J_v (z) = \frac{1}{2 \pi i} \int_L \; w^{-v-1} 
e^{z(w- 1/w)/2} dw,
\label{(17)}
\end{equation}
valid for $Re(z) > 0$.  Correspondence between (\ref{(16a)}) and 
(\ref{(17)}) is 
obtained by letting $w =
(\sqrt{t}
/ \rho)s$, $z = 2\rho \sqrt{t}$, in which case we find for $k = 0$ 
that
\begin{eqnarray}
F_0(t) &\sim& \sum_n \; b_n \left(\frac{\rho}{\sqrt{t}} 
\right)^{n-5/2} \;
\frac{1}{2
\pi i} \int_L
w^{n - 7/2} e^{\rho \sqrt{t}(w - 1/w)} dw\nonumber
\\
&=& \sum_n \; b_n \left( \frac{\rho}{\sqrt{t}} \right)^{n 
- 5/2} J_{5/2 - n} (2\rho
\sqrt{t}).\label{(18)}
\end{eqnarray}
Higher sum-rule moments can be obtained via explicit differentiation 
with respect to $s$ in the
integrand of (\ref{(16a)}); {\it{e.g.}},
\begin{eqnarray}
F_1^p(t) &\sim& \sum_{n=0} \; b_n \left\lbrace \left(\sbody52 - n 
\right) \frac{1}{2 \pi i}
\int_L
e^{st - \rho^2/s} s^{n - 9/2} ds 
- \rho^2 \frac{1}{2 \pi i} \int_L e^{st-\rho^2/s} s^{n- 
11/2} ds \right\rbrace\nonumber
\\
&=& \sum_{n=0} \; b_n \left\lbrace \left(\sbody52 - n 
\right)
\left(\frac{\rho}{\sqrt{t}}
\right)^{n - \frac{7}{2}}\!\! J_{\frac{7}{2} - n} (2 \rho \sqrt{t})
-\rho^2 
\left(\frac{\rho}{\sqrt{t}} \right)^{n - \frac{9}{2}}\!\! J_{\frac{9}{2} - n} 
(2 \rho
\sqrt{t})\right\rbrace.
\label{(19)}
\end{eqnarray}
Finally, we note that Bessel functions of half-integer order can be 
expressed in terms of
elementary trigonometric functions.  We find from (\ref{(18)}) that
\begin{eqnarray}
F_0 (s_0) &=& \frac{3}{4\pi^2 \rho^4} \left \lbrace \sin (2\rho 
\sqrt{s_0}\,) \left[ -\rho^2 s_0 +
\sbody{25}{32} + {\cal O} \left( \frac{1}{\rho^2 s_0} \right) 
\right]\right.\nonumber
\\
& &\qquad\left. 
\quad+ \cos (2 \rho \sqrt{s_0}\,) \left[ - \frac{7\rho s_0^{1/2}}{4} 
+ \frac{15}{128\rho
s_0^{1/2}} + {\cal O} \left(\frac{1}{\rho^3 s_0^{3/2}} \right) \right] 
\right\rbrace \;
.\label{(20)}
\end{eqnarray}
For $\rho^2s_0 > 2$ this approximate expression differs from (\ref{(12)})
with $k = 0$ by
less than 5\%.  Given an instanton size $1/\rho \approx 600\,{\rm MeV}$, eq.
(\ref{(20)}) is seen to oscillate slowly as $s_0$ increases past $1\,\, 
{\rm GeV}^2$, going from positive to
negative as $s_0$ increases past $2.9\,\, {\rm GeV}^2$.  Since the purely-
perturbative contribution is
also
positive and quadratic in $s_0$ [1,2], we see the effect of instanton 
contributions is to enhance
the size of field-theoretic contributions to $F_0^p$ at low $s_0$, 
but to diminish somewhat the
magnitude of field-theoretic contributions for values of the 
continuum threshold chosen to be
above $2.9\,\, {\rm GeV}^2$.  
The corresponding expression for $F_1$ in terms of elementary trigonometric 
functions
can be obtained from  (\ref{(19)}):
\begin{eqnarray}
F_1\left(s_0\right)&=&\frac{3}{8\pi^2\rho^6}\left\{
\sin\left(2\rho\sqrt{s_0}\,\right)\left[-2\rho^4s_0^2+\frac{129}{16}\rho^2s_0
+{\cal O}(1)\right]\right.
\nonumber\\
& &\qquad\qquad +\cos\left(2\rho\sqrt{s_0}\,\right)
\left.\left[
-\frac{11}{2}\rho^3s_0^{3/2}+\frac{531}{64}\rho s_0^{1/2}+{\cal O}
\left(\frac{1}{\rho\sqrt{s_0}}\right)\right]\right\}\quad .
\label{f1}
\end{eqnarray}

Once again, the leading instanton contribution to $F_1$ is seen to be 
lower-degree in $s_0$
than the ${\cal O}\left(s_0^3\right)$ purely-perturbative contribution.  As
a final comment, 
it should be noted that for detailed phenomenological work, all 
the FESRs require inclusion of 
an overall renormalization-group factor which is identical for the 
(leading) perturbative and 
instanton contributions. 

A.S. Deakin, V. Elias and T.G. Steele are grateful for support from 
the Natural Sciences and
Engineering Research Council of Canada.\eject
\begin{flushleft}
{\bf{\hspace{.5cm}References}}
\end{flushleft}
\begin{itemize}
\item[1.] W. Hubschmid and S. Mallik, Nucl. Phys. B 193 (1981) 368; 
T. Truong, Phys. Lett.
B 117 (1982) 109; A.L. Kataev, N.V. Krasnikov, and A.A. Pivovarov, 
Phys. Lett. B 123 (1983)
93.
\item[2.] J. Bijnens, J. Prades, E. de Rafael, Phys. Lett. B 348 
(1995)
226; J. Prades, hep-ph/9708395.
\item[3.]  S.G. Gorishny, A.L. Kataev, and S.A. Larin, Phys. Lett. B 
135 (1984)
457;                
\item[4.]  S.G. Gorishny, A.L. Kataev, S.A. Larin, and L.R. 
Surguladze,
Mod. Phys. Lett. A5 (1990) 2703; K.G. Chetyrkin, Phys. Lett. B 390
(1997) 309.
\item[5.] M.A. Samuel, G. Li, and E. Steinfelds, Phys. Rev. E 51 
(1995)
3911.   
\item[6.] V.A. Novikov, M.A. Shifman, A.I. Vainshtein, and V.I. 
Zakharov, Nucl. Phys. B 191
(1981) 301;  E. Gabrielli and P. Nason, Phys. Lett. { B313}
(1993) 430; P. Nason and M. Poratti, Nucl. Phys. { B421} (1994) 518;
P. Nason and M. Palassini, Nucl. Phys. { B444} (1995) 310.
\item[7.] A.E. Dorokhov, S.V. Esaibegian, N.I. Kochelev, N.G. 
Stefanis, J. Phys. G 23 (1997)
643; E.V. Shuryak, Nucl. Phys. B 214 (1983) 237.
\item[8.] I.S. Gradshteyn and I.M. Ryzhik, Table of integrals, 
series, and products (Academic
Press, New York, 1980) p. 718 [eq. 6.633.3] and p. 1062 [eq.
9.235.2].
\item[9.] {\it{Ibid}}, p. 963 [eq. 8.451.6].
\item[10.] G.F.D. Duff and D. Naylor, Differential equations of 
applied mathematics
(Wiley,
New York, 1966) p. 300.                
\end{itemize}
\eject
\begin{flushleft}
{\bf{\hspace{-.4cm}Figure Captions}}
\end{flushleft}
\begin{enumerate}
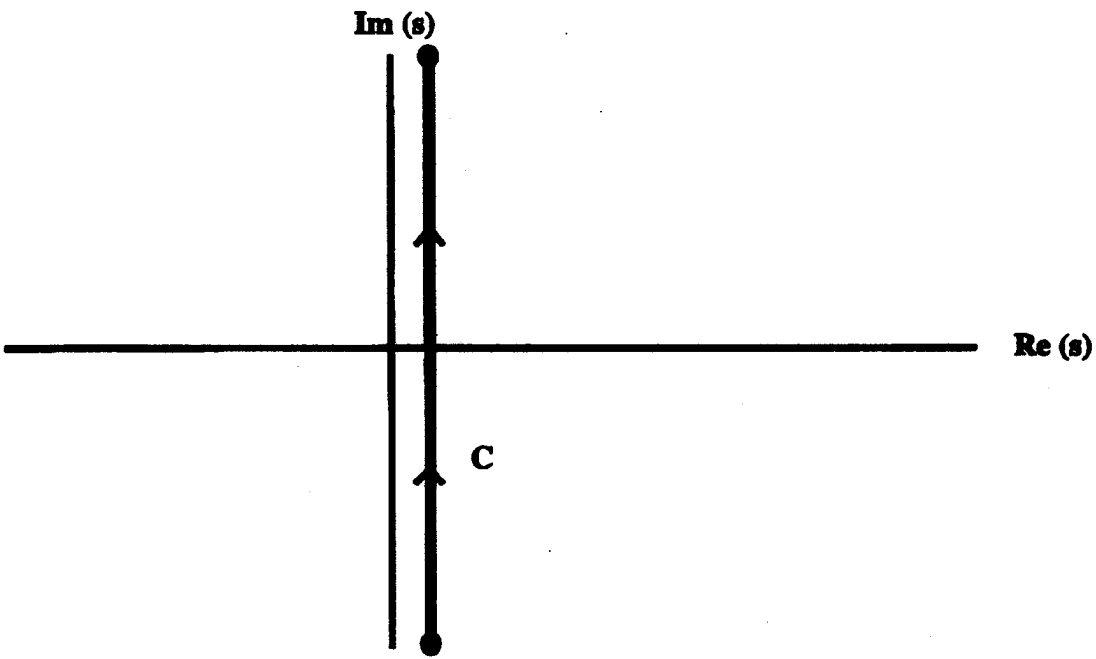
\item[Fig. 1:] The contour $C$ characterizing the inverse Laplace-
transform contour
integral.
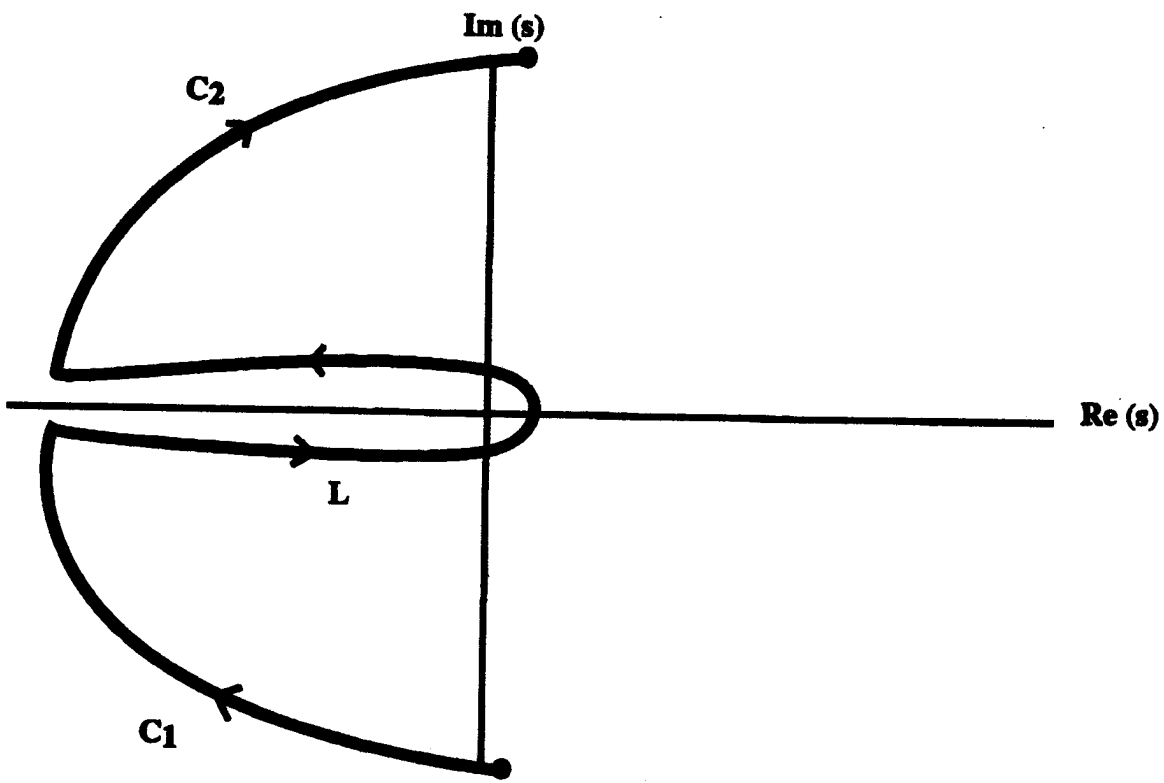
\item[Fig. 2:] Distortion of $C$ into the sum of infinite-radius arc 
contributions $C_{1,2}$ and
the
Hankel loop contour $L$.
\end{enumerate}
\end{document}